\documentstyle[twocolumn,aps,epsfig]{revtex} 
\begin{document}
\twocolumn[\hsize\textwidth\columnwidth\hsize\csname @twocolumnfalse\endcsname

\title{
Thermodynamic properties of the one-dimensional Kondo insulators \\
studied by the density matrix renormalization group method
}

\author{Naokazu Shibata, Beat Ammon$^1$, Matthias Troyer, 
Manfred Sigrist$^2$ and Kazuo Ueda} 
\address{
Institute for Solid State Physics, University of Tokyo,  
7-22-1 Roppongi, Minato-ku, Tokyo 106, Japan \\
$^1$Theoretische Physik, ETH-H\"onggerberg, 8093 Z\"urich, Switzerland\\
$^2$Yukawa Institute for Theoretical Physics, Kyoto University, Kyoto
606-01, Japan}

\date{30 Dec 1997}
\maketitle

\begin{abstract}
Thermodynamic properties of the one-dimensional 
Kondo lattice model at half-filling are studied by the density matrix 
renormalization group method applied to the quantum transfer
matrix. Spin susceptibility, charge 
susceptibility, and specific heat are calculated down to 
$T=0.1t$ for various exchange constants. 
The obtained results clearly show crossover behavior from
the high temperature regime of nearly independent localized spins
and conduction electrons to the low temperature regime where the 
two degrees of freedom couple strongly.
The low temperature energy scales of the charge and spin 
susceptibilities are determined and shown to be 
equal to the quasiparticle gap and the spin gap, 
respectively, for weak exchange couplings.
\end{abstract}

\vskip2pc]
\narrowtext

The Kondo lattice model (KLM) is a simple theoretical model 
for heavy Fermions which
consists of two different types of electrons; the localized 
spins whose charge degrees of freedom are completely suppressed,  
and the conduction electrons that propagate as carriers in extended 
orbitals. 
Both the conduction electrons and the localized spins do not
interact among themselves, however, 
the exchange interaction between them leads to
correlations and yields various interesting physical phenomena.
\cite{PD} 

Recent studies on the one-dimensional KLM has shown that this model 
is always insulating at half-filling.\cite{PD} 
This insulating phase has different excitation gaps 
for the spin and charge channels.
The spin gap defines the energy cost needed for the lowest 
spin excitation which changes the total spin quantum number. 
The charge gap defines that for
the lowest pure charge excitation which changes the 
total carrier number by two keeping the spin quantum numbers fixed.

The spin gap of this insulating phase has been extensively
studied both analytically and numerically.
\cite{sgap1,sgap2,tsvelik,sgap3,fujimoto}
It has been shown that the spin gap $\Delta_s$
has similar $J$-dependence to that of the Kondo temperature 
$T_K$ for the single impurity Kondo model. 
The numerical studies show  
$\Delta_s \propto \exp(-1/\alpha\rho J)$,
where $\rho$ is the density of state
at the fermi level and $\alpha$ is the lattice enhancement
factor.  The lattice enhancement factor was recently determined to be
$\alpha=1.4$ by the density matrix renormalization group (DMRG) 
method.\cite{sgap3}
The similarity between the Kondo temperature and the spin gap 
suggests that the origin of the spin gap 
is due to some sort of singlet formation 
between the conduction electrons and 
the localized spins which is the essence of the Kondo effect.
The difference appears in the coefficient $\alpha$ in the exponent,
which is due to collective singlet formation of conduction electrons 
with many localized spins in the lattice.

For the charge gap, on the other hand, linear $J$-dependence 
($\Delta_c=J/2$) has been obtained for weak exchange couplings.  
\cite{nishino,sgap3} This $J$-dependence may be understood by the 
strong antiferromagnetic correlations among localized spins that
generate a
staggering internal magnetic field on conduction electrons which is
proportional to the exchange constant. It should be noted that the 
correlation length of the spin degrees of freedom is much longer than
the charge correlation length.  Therefore the above arguments are 
justified in spite of the fact that there is no magnetic long range 
order. 

The qualitatively different nature of the spin and charge gaps is 
a consequence of quantum mechanical many body effects among
the conduction electrons and localized spins, and is a
unique feature of this strongly correlated insulating phase.

In the present paper we will study how such interplay between
the conduction electrons and the localized spins appears in 
the temperature dependence of various thermodynamic quantities.
To discuss the thermodynamic properties we use 
the finite temperature density matrix renormalization
group method (finite-$T$ DMRG).
We calculate spin and charge 
susceptibilities and specific heat for various exchange 
constants. We will show clear crossover behavior between 
high and low temperature regimes.  In the high temperature 
regimes the localized spins and the conduction electrons are
only weakly coupled, while in the low temperature regime 
they couple strongly.  Characteristic energy scales in
the low temperature regimes are identified as the spin gap 
and the quasiparticle gap for the spin and charge degrees 
of freedom, respectively.

The Hamiltonian we use here is the one-dimensional KLM 
described by
\begin{equation}
  {\cal H} = -t \sum_{i s} (
  c^{\dag}_{i s} c_{i+1 s} + c^{\dag}_{i+1 s} c_{i s}) + 
  J \sum_i {\bf S}_{{\rm c}i} \cdot {\bf S}_{{\rm f}i}
\end{equation}
where the operator $ c_{is} $ ($ c^{\dag}_{is} $) annihilates
(creates) a conduction electron at site $ i $ with spin $ s $ ($=
\uparrow, \downarrow $) ($ S^{\mu}_{{\rm c}i} = (1 / 2)
\sum_{s,s'} c^{\dag}_{is} \sigma^{\mu}_{ss'} c_{is'} $).
The hopping matrix element is given by $t$ and
$ J $ is the antiferromagnetic exchange coupling between the
conduction electron spins $ {\bf S}_{\rm c} $ and localized spins 
$ {\bf S}_{\rm f} $, both beeing spin 1/2 degrees of freedom.
The density of conduction electrons is 
unity at half-filling.

In order to calculate thermodynamic quantities we use the 
density matrix renormalization group method \cite{DMRG} 
applied to the quantum transfer matrix.
Recently this method of finite-$T$ DMRG was successfully applied to 
the one-dimensional quantum spin systems
to calculate thermodynamic quantities.\cite{FTRG,Wang}
The present study is the first application of the finite-$T$
DMRG to a system with Fermionic degree of freedom.

In the following we briefly outline the method.
We define the transfer matrix 
\begin{eqnarray}
{\cal T}_n(M) & = & 
[e^{-\beta h_{2n-1,2n}/M} e^{-\beta h_{2n,2n+1}/M} ]^M
\end{eqnarray}
where $M$ is the Trotter number.
Here the Hamiltonian $H$ is 
decomposed into two parts $H_{\mbox{\scriptsize odd}}=
\sum_{n=1}^{L/2}h_{2n-1,2n}$ and 
$H_{\mbox{\scriptsize even}}=\sum_{n=1}^{L/2}h_{2n,2n+1}$ such as 
$[h_{2n-1,2n},h_{2n'-1,2n'}]=[h_{2n,2n+1},h_{2n',2n'+1}]=0$.
First we diagonalize the transfer matrix with a small $M$ 
to obtain the maximum eigenvalue $\lambda$ and corresponding 
eigenvector.
Next we calculate generalized density matrix from 
the obtained eigenvector. By diagonalizing the density matrix, 
we chose important basis states which have large eigenvalues for the 
representation of the transfer matrix. 
Using these basis states we increase the Trotter number $M$ 
of the transfer matrix within the fixed number of basis states 
and continue the above procedure until we get
the Trotter number sufficient for a given temperature $T$. 
The free energy of the infinite system is directly obtained from the 
maximum eigenvalue $\lambda$ of the transfer matrix: 
$F=-(T/2) \ln \lambda $. 
The spin and charge susceptibilities, and specific heat are
obtained by numerical derivatives of the free energy.

The following calculations are performed by the 
infinite system algorithm
of the finite-$T$ DMRG keeping 40 states per blocks.\cite{FTRG,Wang}
The truncation error in the finite-$T$ DMRG calculation is 
typically $10^{-3}$ and $10^{-2}$ at the lowest temperature
with the Trotter number $M=50$. 
To check the results we have compared the 
obtained spin and charge susceptibilities to those 
calculated by the Quantum Monte Carlo simulations 
for $J/t=1.6$.\cite{Fye}
The overall structure agrees well, but low temperature 
part of our results are more reliable. For example,
the estimated spin gap energy $0.7t$ in
Ref.~11 is not consistent with $0.4t$ 
obtained by the standard zero-temperature DMRG\cite{sgap3},
while it is consistent with the present result 
$(0.45 \pm 0.1)t$ obtained by the finite-$T$ DMRG.

We first show the temperature dependence of the uniform
spin susceptibility. 
The spin susceptibility is obtained from the change 
of the free energy by a small magnetic field $h$: 
$\delta F=\chi_s h^2/2$.
The results for $J/t=0, 1.0, 1.2, 1.6$ and $2.4$ are shown 
in Fig.~1. 
In the limit of $J/t=0$, both the localized spins and 
conduction electrons are uncorrelated.
The susceptibility is given by the sum of the Curie term 
of localized spins and the Pauli term of free 
conduction electrons. 
In this system the contribution of the Pauli 
susceptibility of the free electrons (shown in Fig.~1)
is relatively small, and the total susceptibility of $J/t=0$ 
is dominated by the Curie term of the localized spins, 
which diverges in the limit of $T=0$.

With introducing the exchange coupling, 
low temperature part of $\chi_s$ sharply
decreases with decreasing the temperature.
This drastic change is due to the formation 
of the spin singlet state between the localized 
spins and conduction electrons whose energy scale 
is given by the spin gap
for small exchange constant.
The spin gap $\Delta_s$ obtained by the zero temperature 
DMRG is $0.08t$ for $J=1.0t$, which is consistent with
the characteristic temperature at which $\chi_s$ 
starts to decrease. 

\begin{figure}
  \epsfxsize=85mm \epsffile{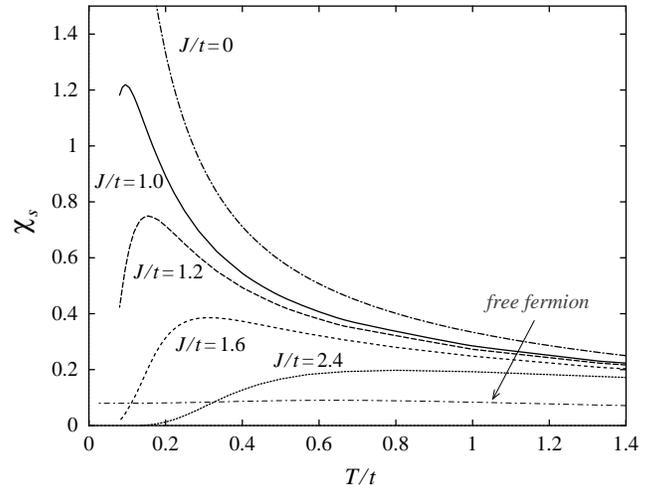}
\caption[*]{Spin susceptibility of the half-filled 
one-dimensional Kondo lattice model. The truncation error 
in the finite-$T$ DMRG calculation is 
typically $10^{-3}$ and $10^{-2}$ at the lowest temperature.
}
\label{sus_all}
\end{figure}

This characteristic temperature 
separates high temperature region where 
$\chi_s$ increases with decreasing the 
temperature, and low temperature region where 
the susceptibility drops rapidly with 
decreasing temperature.
This temperature is expected to be 
determined by the competition between the thermal 
fluctuations and the singlet correlations.
Since the exchange interaction stabilize the singlet 
correlation between the localized spins and 
conduction electrons, the crossover temperature 
increases with enhanced exchange coupling.

In order to determine the energy scale at low temperatures 
more precisely we estimate the activation energy by fitting the 
obtained susceptibility. 
The estimated activation energy for the spin susceptibility
is summarized in Table I for $J/t=1.6t, 2.4t$, and $ 3.0t$. 
Here we can compare this energy scale with 
quasiparticle gap and spin gap, both responsible 
for the magnetic excitation.
The quasiparticle gap and spin gap are 
obtained by the standard zero temperature 
DMRG method.

In Table I we see that the lower
one of the quasiparticle gap and the spin gap determines 
the low temperature energy scale of the spin 
susceptibility.
This is consistent with the general form of
the susceptibility that is formally written as 
\begin{eqnarray} 
    \chi_s &=& Z^{-1} N^{-1} \beta \sum_{m} {\rm e}^{-\beta E_m}
    {\langle m | S_z^{\rm total} | m \rangle} ^2\  ,\\
       Z &=& \sum_m {\rm e}^{-\beta E_m}.
\end{eqnarray}
According to the above equation the low temperature 
behavior is determined by the energy difference 
between the ground state and the lowest excited 
state which is active for magnetic field.

Next we consider the charge susceptibility.
The charge susceptibility $\chi_c$ is obtained from the change 
of free energy by a small shift of chemical potential $\mu$,
$\delta F=\chi_c \mu^2/2$. In the present calculation
we use the fact that 
the chemical potential is zero at half-filling 
owing to the SO(4) symmetry of the model.\cite{nishino}

Calculated $\chi_c$ for $J/t=0, 1.0, 1.2, 1.6$ and $2.4$ 
are shown in Fig.~2. 
For $J/t=0$, $\chi_c$ does not show diverging 
behavior at low temperatures in contrast to $\chi_s$. 
In the limit of $T=0$ $\chi_c$ is equal to
the density of state of conduction electrons
that is $1/\pi$.
This is natural since the charge degrees of freedom is 
solely governed by the conduction electrons.
Since there is no correlation between up-spin 
and down-spin conduction electrons for $J/t=0$, 
$\chi_c/4$ is equal to the spin susceptibility 
of the free conduction electrons.
The slight increase in $\chi_c$ in the low temperature
region is a characteristic feature of the
one-dimensional system where the density of states 
diverges at the band edge. 

\begin{figure}
  \epsfxsize=85mm \epsffile{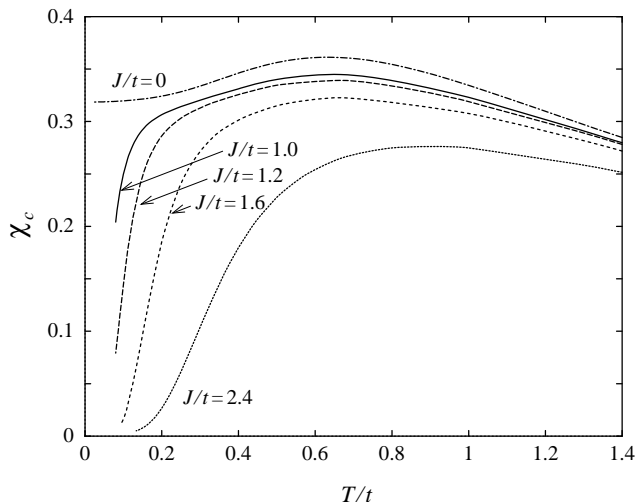}
\caption[*]{Charge susceptibility of the half-filled 
one-dimensional Kondo lattice model. 
}
\label{Csus}
\end{figure}

Switching on the exchange coupling, 
$\chi_c$ decreases rapidly at low temperatures. 
Similarly to $\chi_s$, this is due to the singlet 
formation between the conduction electrons and the localized spins. 
With increasing the exchange coupling,
decrease in $\chi_c$ appears at higher temperature.
This is consistent with the behavior of the charge gap
(quasiparticle gap), which is enhanced with increasing the 
exchange coupling.

To determine the energy scale at low temperatures, 
we estimate activation energy for the charge susceptibility.
By fitting $\chi_c$ with an exponential function,
activation energy $\Delta_{\chi_c}$ is obtained 
as listed in Table I for $J/t=1.6t, 2.4t$, and $ 3.0t$.
From this table it is concluded that the quasiparticle gap always
determines the low temperature energy scale of the charge
susceptibility. Since the charge gap is twice the
quasiparticle gap, the lowest excitation that is
responsible for the charge excitations is always the quasiparticle 
gap. 

Finally, we consider the specific heat.
The specific heat is calculated from the second derivative 
of the free energy; $C=-T\partial^2 F/\partial T^2$.
The results for $J/t=0, 1.0, 1.2, 1.6$ and $2.4$ are shown in Fig.~3.

At $J/t=0$ the specific heat of this model is given by the 
sum of the delta function at $T=0$ that originates from the localized 
spins and the specific heat of free conduction electrons.
By including the exchange coupling, they are combined 
to make a two-peak structure. 
The peak at higher temperatures is almost independent of 
the exchange constant, and similar to the structure of the free 
conduction electrons. Thus, this structure originates 
from the band structure effect of the conduction electrons.
The peak at lower temperatures strongly depends on the exchange 
constant. The peak shifts toward higher temperatures and becomes 
broader with increasing $J/t$.
This behavior is consistent with that of
the spin susceptibility whose peak also shifts 
toward higher temperatures and becomes broader with increasing $J/t$.
Thus, we can assign the origin of the peak at lower temperatures 
due to the localized spin degrees of freedom coupled through the
conduction electron spins.
The slight decrease in the specific heat at low temperatures 
for weak $J/t$ compared with that of the free conduction 
electrons is due to the opening of the charge gap.
The decrease at low temperatures are compensated 
by the increase at higher temperatures as shown in Fig.~3.

Further increasing the exchange coupling, large
excitation gap which is comparable to the hopping
matrix element $t$ opens for both spin and charge sectors.
The ground state is close to the collection of the local singlets,
and the specific heat has a single peak structure.
(See J/t=2.4 in Fig.~3)

\begin{table}
\caption{Activation energy obtained from the 
spin and charge susceptibility, $\Delta_{\chi_s}$ and 
$\Delta_{\chi_c}$, and the quasiparticle gap $\Delta_{qp}$ and 
the spin gap $\Delta_s$ of the one-dimensional Kondo lattice model.
The charge gap is twice the quasiparticle gap;
$\Delta_c = 2\Delta_{qp}$ }
\begin{tabular}{l|cccc} 
 & $\Delta_{\chi_s}$ & $\Delta_{\chi_c}$ & $\Delta_s$ 
& $\Delta_{qp}$ \\ \hline
$J=1.6t$ & $0.45t \pm 0.1t$& $0.6t \pm 0.1t$ & $0.4t $ & $ 0.7t$  \\
$J=2.4t$ & $1.2t \pm 0.1t$ & $1.0t \pm 0.1t$ & $1.1t $ & $ 1.1t$  \\
$J=3.0t$ & $1.6t \pm 0.1t$ & $1.4t \pm 0.1t$ & $1.8t $ & $ 1.5t$  \\ 
\end{tabular}
\label{table-1}
\end{table}

\begin{figure}
  \epsfxsize=85mm \epsffile{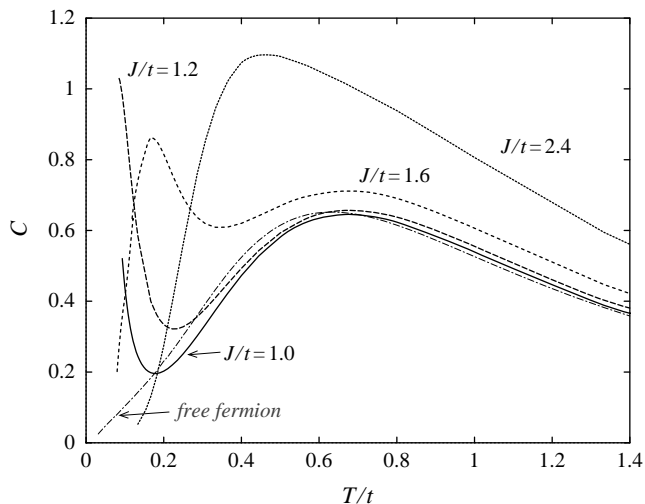}
\caption[*]{Specific heat of the half-filled 
one-dimensional Kondo lattice model. }
\label{Spec}
\end{figure}

In conclusion we have successfully applied the finite-$T$ 
DMRG to the one-dimensional Kondo lattice model at half-filling.
Temperature dependence of the spin and charge susceptibility 
as well as the specific heat are calculated down to $T=0.1t$.
Compared with the quantum Monte Carlo simulations, an 
advantage of the finite-$T$ DMRG is that it is free from 
statistical errors and numerical errors in the calculations
are under control to certain extent by keeping sufficient number of 
basis states.  More importantly it is, in principle, 
free from the negative sign 
problem which practically invalidates the quantum Monte Carlo
simulations for frustrated spin systems and most of Fermionic 
systems.  

The subtle interplay between the localized spins and the 
conduction electrons are clearly seen in the temperature 
dependence of the susceptibilities and the specific heat.
The low-temperature energy scale of the spin susceptibility 
is determined by the smaller one of the spin gap and the 
quasiparticle gap.  Thus, the spin gap is the low energy 
scale in the weak coupling regime.  On the other hand, the
low energy scale for the charge susceptibility is always
determined by the quasiparticle gap, which is half of the
charge gap. Effects of both the spin gap and the charge gap 
are seen in the specific heat.
 
The present study is the first one where the finite-$T$ 
DMRG is applied to a system with Fermions.  Existence of the
excitation gap at half-filling is favorable for convergence of
numerical calculations. Of course it is an interesting 
future problem to apply the finite-$T$ DMRG to the 
Kondo lattice model away half-filling where the ground state is 
a Tomonaga-Luttinger liquid with gapless spin and charge
excitations\cite{TL1,TL2}.

We would like to thank T.~Nishino for helpful advice and
valuable discussion.  This work is financially supported by
a Grant-in-Aid from the Ministry of Education, Science and 
Culture. N.~S. is supported by the Japan Society for the 
Promotion of Science.

\end{document}